\documentclass[final,5p,times,twocolumn]{elsarticle}

\usepackage{amssymb}
\usepackage{graphicx}
\usepackage{dcolumn}
\usepackage{bm}
\usepackage{amsmath}
\usepackage{pbox}

\newcommand{\multilinebox}[1]{\pbox{\linewidth}{\vspace{.5\baselineskip}#1\vspace{.5\baselineskip}}}

\journal{Physics Letters B}

\begin{document}

\begin{frontmatter}



\title{From neutron skins and neutron matter to the neutron star crust}


\author[inst1]{William G. Newton}

\affiliation[inst1]{organization={Department of Physics and Astronomy},
            addressline={Texas A\&M University-Commerce}, 
            city={Commerce},
            postcode={75429}, 
            state={TX},
            country={USA}}

\author[inst1]{Rebecca Preston}
\author[inst1]{Lauren Balliet}
\author[inst1]{Michael Ross}


\begin{abstract}
We present the first Bayesian inference of neutron star crust properties to incorporate neutron skin data, including the recent PREX measurement of the neutron skin of $^{208}$Pb, combined with recent chiral effective field theory predictions of pure neutron matter with statistical errors. Using a compressible liquid drop model with an extended Skyrme energy-density functional, we obtain the most stringent constraints to date on the transition pressure $P_{\rm cc}=0.33^{+0.07}_{-0.07}$ MeV fm$^{-3}$ and chemical potential $\mu_{\rm cc}=12.6^{+1.8}_{-1.9}$ MeV (which control the mass, moment of inertia and thickness of a neutron star crust), the proton fractions that bracket the pasta phases $y_{\rm p}=0.115^{+0.016}_{-0.017}$ and $y_{\rm cc}=0.041^{+0.007}_{-0.006}$, as well as the relative mass and moment of inertia $\Delta M_{\rm p} / \Delta M_{\rm c}\approx \Delta I_{\rm p} / \Delta I_{\rm c} = 0.54^{+0.05}_{-0.09}$ and thickness $\Delta R_{\rm p} / \Delta R_{\rm c}=0.129^{+0.019}_{-0.030}$ of the layers of non-spherical nuclei (nuclear pasta) in the crust.
\end{abstract}



\end{frontmatter}


\section{Introduction}
\label{sec:intro}

In the deepest layers of a neutron star crust, nuclear sizes and separations become comparable. Coulomb and surface energy competition predict that nuclei deform and fuse through a sequence of shapes termed nuclear pasta that cushion the base of the crust \cite{Ravenhall:1983fk,Hashimoto:1984qy,Rea:2015uq}. The extent of the crust and pasta phases therein could have observable consequences \cite{Newton:2014pb}, affecting crustal oscillation modes \citep{Steiner:2009wo,Gearheart:2011tg,Sotani:2012oz}, crust cooling \citep{Brown:2009aa,Horowitz:2015rt,Newton:2013dz}, crust shattering \citep{Tsang:2012aa,Neill:2021tg}, persistent gravitational waves from mountains \cite{Gearheart:2011tg}, magnetic field evolution \citep{Pons:2013ly} and damping of core modes \citep{Wen:2012aa,Vidana:2012aa}. In order to make progress understanding these phenomena, it is important to constrain the size of the crust and the amount of pasta therein. 

The compressible liquid drop model (CLDM) \cite{Baym:1971aa,Lattimer:1991yq,Watanabe:2000vn,Schneider:2017aa} is an efficient method of calculating crustal composition, mass and thickness. A Wigner-Seitz cell containing one nucleus or segment of nuclear pasta surrounded by the neutron gas is modeled. Theoretical uncertainty arises on two fronts: the pure neutron matter (PNM) equation of state (EOS) that governs the hydrostatic pressure and chemical potential in the deep layers of the crust, and the interface energy of the nuclei and pasta shapes, which is specified separately to the bulk EOS \cite{Gearheart:2011tg,Carreau:2019aa,Balliet:2021lr,McNeil-Forbes:2019wx,Dinh-Thi:2021lr,Grams:2021fj,Grams:2021kx}. 

In this letter, we perform, for the first time using neutron skin data, a Bayesian inference of the crust-core transition pressure $P_{\rm cc}$ which controls the mass of the crust, crust-core transition baryon chemical potential $\mu_{\rm cc}$ which controls the thickness of the crust, the thickness of the pasta phases relative to the crust $\Delta R_{\rm p} / \Delta R_{\rm c}$, the relative mass and moment of inertia of the pasta phases $\Delta M_{\rm p} / \Delta M_{\rm c} \approx \Delta I_{\rm p} / \Delta I_{\rm c}$, and the proton fraction at the top ($y_{\rm p}$) and bottom ($y_{\rm cc}$) of the pasta phases. We combine this with theoretical PNM calculations \cite{Drischler:2020ab} with well quantified errors. Neutron skin measurements and PNM calculations are known to inform the location of the crust-core boundary and the amount of nuclear pasta \cite{Horowitz:2001aa,Fattoyev:2012uo,Tews:2017zl,Balliet:2021lr,Dinh-Thi:2021lr}.  We account for all uncertainties in the CLDM model used, and detail the remaining sources of uncertainty still to be addressed.


%
%

\section{Model}
\label{sec:model}

Neutron skins, PNM, and bulk nuclear matter in the CLDM are modeled using the extended Skyrme energy-density functional (EDF) \cite{Zhang:2016ww,Lim:2017aa} which allows us to explore a wide range of density dependences of the EOS in the range important for nuclear pasta, the crust-core transition and finite nuclei: 0.25  - 1 $n_0$ where $n_0=0.16$ fm$^{-3}$ is nuclear saturation density. 

The full expression for the extended Skyrme energy density functional (EDF) used here can be found in \cite{Newton:2020aa}. The parts of the Skyrme EDF $\mathcal{H}$ that we use to control the symmetry energy are the zero-range and density-dependent terms

\begin{equation}
    \mathcal{H}_{\rm \delta} =  \frac{1}{4} t_{0} \rho^2 [(2+x_{0}) - (2x_{0}+1)(y_{p}^{2}+y_{n}^{2})],
\end{equation}

\begin{align}
    \mathcal{H}_{\rho} &=  \frac{1}{4} t_{3} \rho^{2+\alpha_3} [(2+x_{3}) - (2x_{3}+1)(y_{p}^{2}+y_{n}^{2})] \notag \\
    &+  \frac{1}{4} t_{4} \rho^{2+\alpha_4} [(2+x_{4}) - (2x_{4}+1)(y_{p}^{2}+y_{n}^{2})]
\end{align}

\noindent where $y_{\rm p}$ and $y_{\rm n}$ are the proton and neutron fractions and $\alpha_3$, $\alpha_4$, $x_3, x_4$, $t_3$ and $t_4$ are parameters. $x_0$, $x_3$ and $x_4$ can be adjusted to control the density dependence of the PNM EOS, while the remaining parameters can be readjusted to maintain symmetric nuclear matter properties at the values of the baseline Skyrme Sk$\chi$450 \cite{Lim:2017aa} within the accepted values from nuclear experiment.

The interface tension, consisting of surface $\sigma_{\rm s}(y_{\rm p})$ and curvature $\sigma_{\rm c}(y_{\rm p})$ terms, is given as a function of proton fraction $y_{\rm p}$ \cite{Ravenhall:1983uq,Lattimer:1985fj,Lorenz:1991thesis}:

\begin{equation} \label{eqn:surf}
    \sigma_{\rm s}(y_{\rm p})= \sigma_0 \frac{ 2^{p+1} + b}{ \frac{1}{y_{\rm p}^p} + b + \frac{1}{(1-y_{\rm p})^p} }; \;\;\; \sigma_{\rm c}(y_{\rm p}) = \sigma_{\rm s}(y_{\rm p}) \frac{\sigma_{\rm 0,c}}{\sigma_{0}},
\end{equation}

\noindent where the following four model parameters are identified: $\sigma_0$ and $\sigma_{0,c}$ control the strength of the surface and curvature tension in symmetric matter $y_{\rm p}=0.5$ and $p$ and $b$ control the isospin dependence. In particular, the parameter $p$ controls the behavior of the surface energy in very neutron rich environments $y_{\rm p}\to 0$ \cite{Newton:2013sp,Carreau:2019aa}, and is particularly important in the modeling of the crust. The curvature tension has been extended in some works to include its own proton fraction dependence \cite{Newton:2013sp,Carreau:2019aa} with additional model parameters, which we do not use here.


%
%

\subsection{Model parameters}
\label{sec:modpar}

The nuclear symmetry energy $S(n)$, a function encoding the energy cost of replacing protons with neutrons in a nuclear system, is a convenient intermediary between nuclear and neutron star properties. It is defined in the expansion of the energy per particle of uniform nuclear matter about a proton fraction of one half. Similarly one can define an expression for the surface symmetry tension $\sigma_{\delta}$, so we have:

\begin{align} \label{eq1} &E(y_{\rm p},n)=E_0(n) +  S(n) \delta^2 + \dots ; \notag \\
&\sigma(y_{\rm p}) = \sigma_0 + \sigma_{\delta}(p,c) \delta^2 + \dots, \end{align} 

\noindent where $\delta=1-2y_{\rm p}$. The symmetry energy can then be expanded as a power series in density about nuclear saturation density $n_0=0.16$ fm$^{-3}$: using the parameter $\chi=(n-n_0)/3n_0$,  

\begin{equation} \label{eq2} 
S(n) = J + \chi L + \chi^2 K_{\rm sym} + \dots, \end{equation}

The strong correlations between the magnitude, slope and curvature of the symmetry energy $J$, $L$ and $K_{\rm sym}$ and neutron star properties such as their radii and proton fraction, have led to a sustained experimental effort to infer them from nuclear observables \cite{Tsang:2012qy,Li:2014fj,Lattimer:2014uq,Horowitz:2014aa,Tsang:2019ab,Lynch:2021rt}.

$J$, $L$ and $K_{\rm sym}$ can be written in terms of the parameters $x_0$, $x_3$ and $x_4$ in the extended Skyrme model \cite{Newton:2020aa} allowing us to explore the symmetry energy parameter space by translating ranges of $J$, $L$ and $K_{\rm sym}$ into Skyrme models, keeping all symmetric nuclear matter parameters fixed. 


To relate the parameters of our surface energy model Eqn~\ref{eqn:surf} to its isospin expansion Eqn~\ref{eq1} we note that the surface symmetry tension is related to the model parameters via \cite{Newton:2013sp}

\begin{equation} \label{eqn:sdelta}
    \sigma_{\delta} = \sigma_0 \frac{2^p p(p+1)}{2^{p+1} + b}.
\end{equation}

Relating the model parameters to the surface symmetry energy allows us to include a generic correlation between the surface and bulk symmetry energies that emerges from
nuclear mass fits and semi-infinite nuclear matter (SINM) calculations \cite{Danielewicz2003a,Steiner:2005vn} which can be written \cite{Newton:2013sp}

\begin{equation} \label{eq:sigJ} \sigma_{\delta}(J,c) = \frac{J n_{\rm s}^{2/3}}{ (36 \pi)^{1/3}} [ (0.046 \; {\rm MeV}^{-1} c + 0.01\; {\rm MeV}^{-1} ) J - c ]. \end{equation}

\noindent $c$ controls the slope of the correlation \cite{Steiner:2005vn,Newton:2013sp} which depends on the exact methods used to extract the surface energy of which there are a variety \cite{Danielewicz2003a,Steiner:2005vn}.
We take the parameter $c$ to vary widely over a range that encloses all relations extracted from empirical and theoretical fits.

We thus replace the model parameter $b$ in~Eqn(2) with the parameter $c$ defined above; given a value of $c$, $b$ can then be calculated combining~Eqns (5) and (6).

It is important to note that we used the full bulk energy density functional (the extended Skyrme) and surface energy functional (Eqn.~\ref{eqn:surf}) in our calculations, and not the truncated expansions \ref{eq1}-\ref{eq2} which provide us convenient model parameters.

To summarize, we have seven model parameters: $J$, $L$ and $K_{\rm sym}$ controlling the bulk EOS and $\sigma_0$, $\sigma_0,c$ and $p$ controlling the surface energy.

\begin{figure}[!t]
\centering
    \begin{tabular}{c|c|c|c|c}
        \hline
		Prior & \hspace{-0.1cm}\multilinebox{\;\;\;\;\;\; $\sigma_0$ \\(MeV fm$^{-2})$}\hspace{-0.1cm} & $p$ & $c$ & \hspace{-0.1cm}\multilinebox{\;\;\;\;\;\; $\sigma_{0,c}$ \\(MeV fm$^{-2})$}\hspace{-0.1cm}\\ 
		\hline
		Surf:Unif  & 0.8-1.3 & 2.0-4.0  & 2.0-7.0  & 0.0-1.0 \\ 
		Surf:Fit & 1.1 & 3.8$\pm$0.02 & 4.5 & 0.6 \\ 
		\hline
    \end{tabular}
\centerline{\includegraphics[scale=0.28]{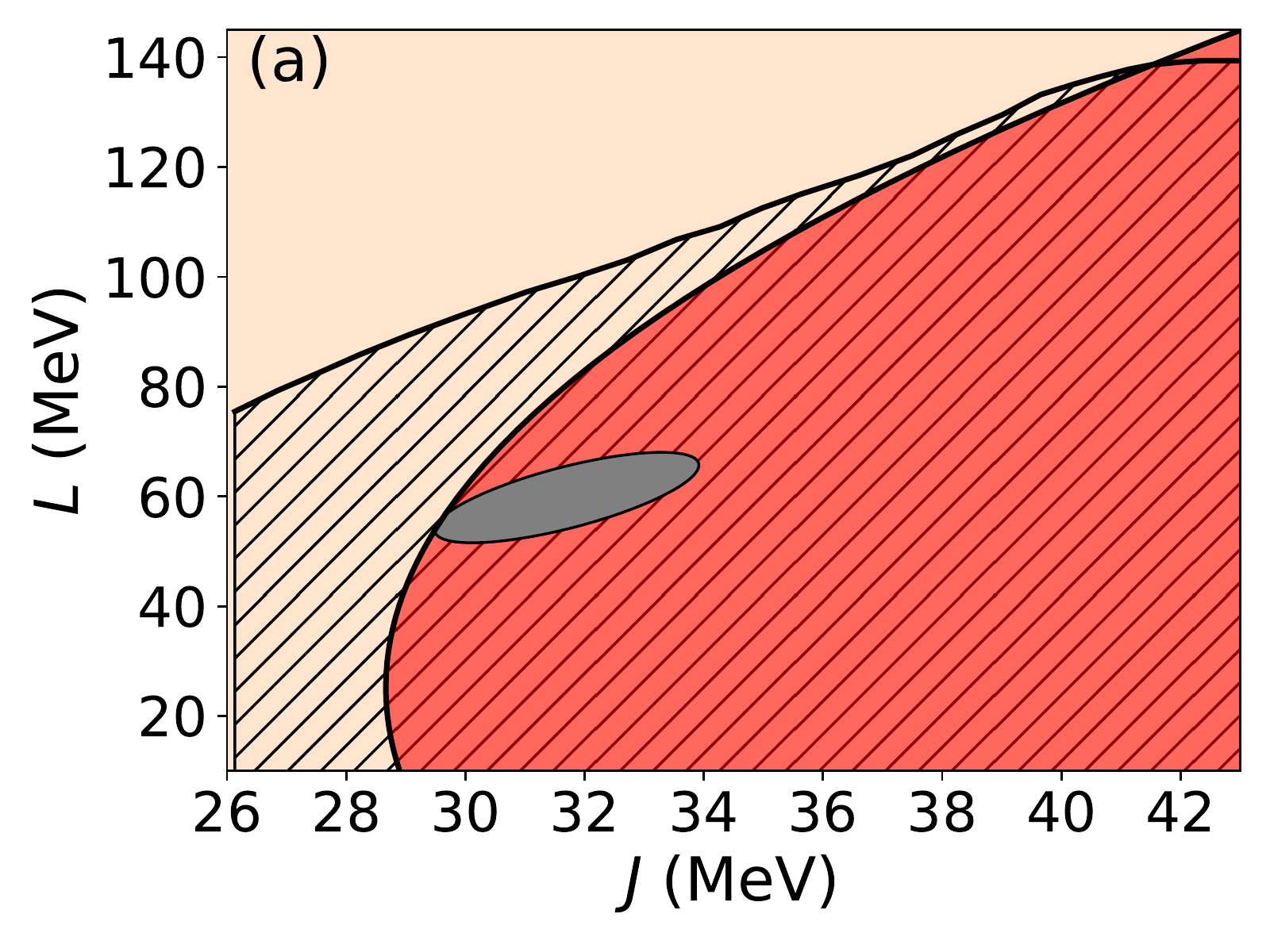}\includegraphics[scale=0.28]{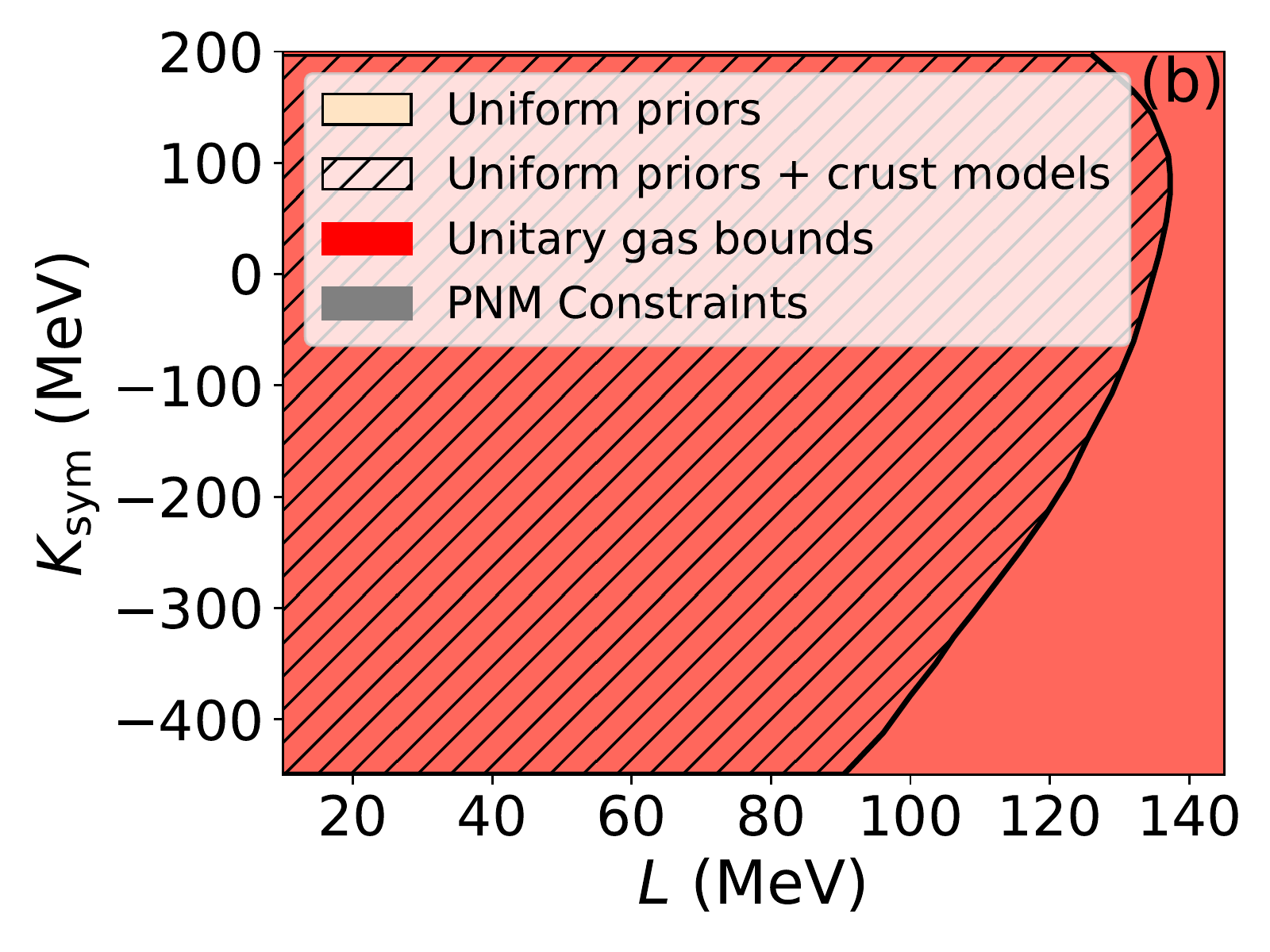}}
\caption{Top: Table of surface parameter ranges and values corresponding to the two surface priors used. Bottom: Symmetry energy priors used, shown in $J-L$ space (a) and $L-K_{\rm sym}$ space (b). The whole space shown (light peach) is the uniform prior we start with. The regions for which viable crust models exist are shown as the hatched region, and the region consistent with unitary gas constraints in red. The uniform and unitary gas constraint coincide in $L$-$K_{\rm sym}$ space since the latter does not constrain $K_{\rm sym}$ \cite{Tews:2017zl}. The $1-\sigma$ error ellipse from chiral-EFT calculations \cite{Drischler:2020aa} is shown in grey (a).} \label{fig:priors}
\end{figure}


\begin{figure*}[!t]
\centerline{\includegraphics[scale=0.3]{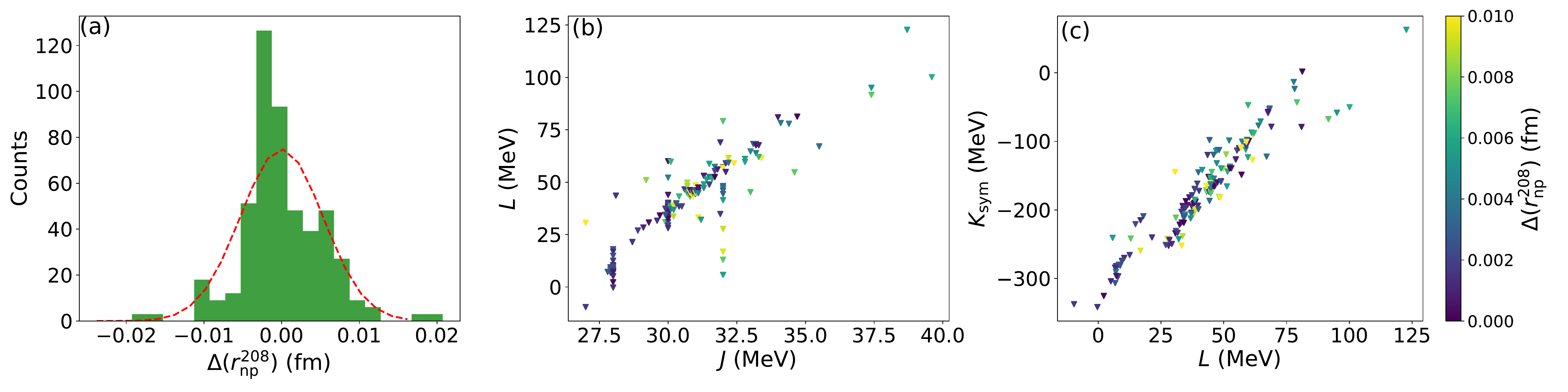}}
\caption{A histogram (a) of the differences $\Delta(r^{208}_{\rm np})$ between the neutron skins of $^{208}$Pb predicted by  the 91 of the most recent Skyrmes from \cite{Dutra:2012wd}, and extended Skyrmes that have been refit to each value of $J$,$L$ and $K_{\rm sym}$ from that set. We show the values of $J$ versus $L$ (b) and $L$ versus $K_{\rm sym}$ (c) common to two sets of Skyrme parameters: The colors indicate $\Delta(r^{208}_{\rm np})$. There is no obvious correlation between each symmetry energy parameter and the difference between the neutron skin predictions, so we may treat $\Delta(r^{208}_{\rm np})$ as a random variable. The distribution in (a) can be mimicked by a Gaussian of standard deviation 0.0045fm.}
\end{figure*}

%
%

\subsection{Priors}
\label{sec:priors}

Priors are given in Fig.~1. We start with uniform priors on both the symmetry energy and surface parameters, over a wide enough range to incorporate all experimental constraints. Not all of the parameter space gives viable crust models: in particular, the large $L$, small $J$ EOSs lead to PNM instabilities at crust densities. Although these models are effectively filtered out, we still refer to the remaining hatched region of Fig.~1 as our uniform prior and label it \emph{Sym:Unif}. We can additionally include theoretical constraints from the unitary gas limit of PNM at very low densities \cite{Schwenk:2005vn,Tews:2017zl,Zhang:2017rm,Kievsky:2018tn} leaving the red region in Fig.~1 (\emph{Sym: Unif+Unit} prior). Note that in $L-K_{\rm sym}$ space the uniform and unitary gas priors coincide.

We refer to the uniform surface priors with the label \emph{Surf:Unif}. These cover a range within which determinations of surface parameters from mass fits and SINM calculations reside. As a more restrictive set of surface priors we use a fit of the surface parameters to three-dimensional Hartree-Fock (3DHF) calculations of crust nuclei near the onset of nuclear pasta  $n\sim0.25n_0$ (\emph{Surf:Fit}) \cite{Balliet:2021lr}. Such calculations are computationally intensive and we can only perform a limited number of them, and so it is difficult to obtain robust model uncertainties on the best fit values of most of the parameters: the fits are most sensitive to the parameter $p$ and that is the only parameter we can currently report an error on; we fix the other parameters at their best fit value. The main purpose of these priors is to provide a way to quantify the effect of our lack of knowledge about the surface parameters.

%
%

\subsection{Data}
\label{sec:data}

Model parameters and crust parameters are inferred using the following data (given with labels used in this text). A recent measurement of the parity violating weak asymmetry in electron scattering off of $^{208}$Pb from the PREX collaboration: $r^{208}_{\rm np} = 0.283\pm0.071$ \cite{Adhikari:2021tm} (this dataset is labeled `PREX'); all available neutron skin measurements of $^{48}$Ca, $^{132}$Sn and $^{208}$Pb (including PREX) combined in quadrature as detailed in \cite{Newton:2020aa}, in addition to measurements of $r^{48}_{\rm np}$ from proton scattering \cite{Shlomo:1979aa,Clark:2003aa}, pion scattering \cite{Gibbs:1992aa}, pionic atoms \cite{Friedman:2012aa} and alpha scattering \cite{Gils:1984aa}: $r_{\rm np}^{48} =0.14\pm0.015$ fm, $r_{\rm np}^{132} =0.24\pm0.04$ fm and $r_{\rm np}^{208} = 0.178\pm0.011$ fm  (labeled `Skins'); a Bayesian inference of $J$ and $L$ from chiral-EFT calculations of PNM, whose 68\% credible region can be represented by a Gaussian distribution with means $J=31.7$, $L=59.8$, variances of $\sigma_J^2=1.11^2$, $\sigma_L^2=4.12^2$ and a covariance of $\sigma_{JL}=3.27$ \cite{Drischler:2020ab}, a region shown in Fig. 1a as the grey ellipse (labeled `PNM'). We also use the data combinations (`PNM+PREX') and (`PNM+Skins').

%
%

\subsection{Modeling neutron skins}
\label{sec:nskinunc}

We use ensembles of extended Skyrme energy-density functionals (EDFs) in 1D Hartree-Fock calculations of neutron skins using the Skyrme RPA code \cite{Colo:2013bh} to infer posteriors of $J$,$L$ and $K_{\rm sym}$ from the neutron skin and PNM data; the Bayesian inference is detailed in full in \cite{Newton:2021dq}.

Our ensembles of Skyrme parameterizations are created by varying $x_0$,$x_3$, and $x_4$ while holding the other parameters constant. However, this neglects correlations between the symmetry energy parameters and other nuclear matter parameters that would arise if the whole functional was fit to, for example, mass data for each choice of $x_0$,$x_3$, and $x_4$, something that is beyond the scope of this study but is the subject of ongoing work. These other nuclear matter parameters - for example the isovector gradient coefficient $G_{\rm v}$ - might also influence the neutron skin. There is evidence that, in the case of neutron skins, the error this introduces is small; for example, the isoscalar and isovector gradient coefficients show little correlation with the neutron skins \cite{Chen:2010aa,Zhang:2014vd}.

In order to assess this source of modeling uncertainty, we calculate $r_{\rm np}^{{208}}$ for 91 different Skyrme interactions created in the past 20 years taken from \cite{Dutra:2012wd}, fit to various subsets of nuclear data. We then construct 91 extended Skyrme interactions with $x_0$,$x_3$, and $x_4$ adjusted to produce the value of $J$, $L$ and $K_{\rm sym}$ of each of the 91 original Skyrmes. We calculate the difference in $r_{\rm np}^{208}$ for each pair of Skyrmes - the original and its corresponding extended Skyrme counterpart - and plot the results in Fig~2. Figs~2a and~2b show the distribution of symmetry energy parameters for the models, with the differences in $r_{\rm np}^{208}$ between the original Skyrmes and the equivalent extended Skyrmes indicated by the color, from lower differences (blue) to higher (yellow). There is no evidence of correlation between the difference in $r_{\rm np}^{208}$ and the values of $J$, $L$ and $K_{\rm sym}$, so we model the errors as random. In Fig~1c we show the distribution of $\Delta(r_{\rm np}^{208})$. The distribution can be reasonable modeled by a Gaussian of standard deviation 0.0045 fm. We thus account for this uncertainty by adding an extra $\pm0.0045$fm error to our neutron skin data points.


\begin{figure}[!t]
\vspace{-1.0cm}
\centerline{\includegraphics[scale=0.4]{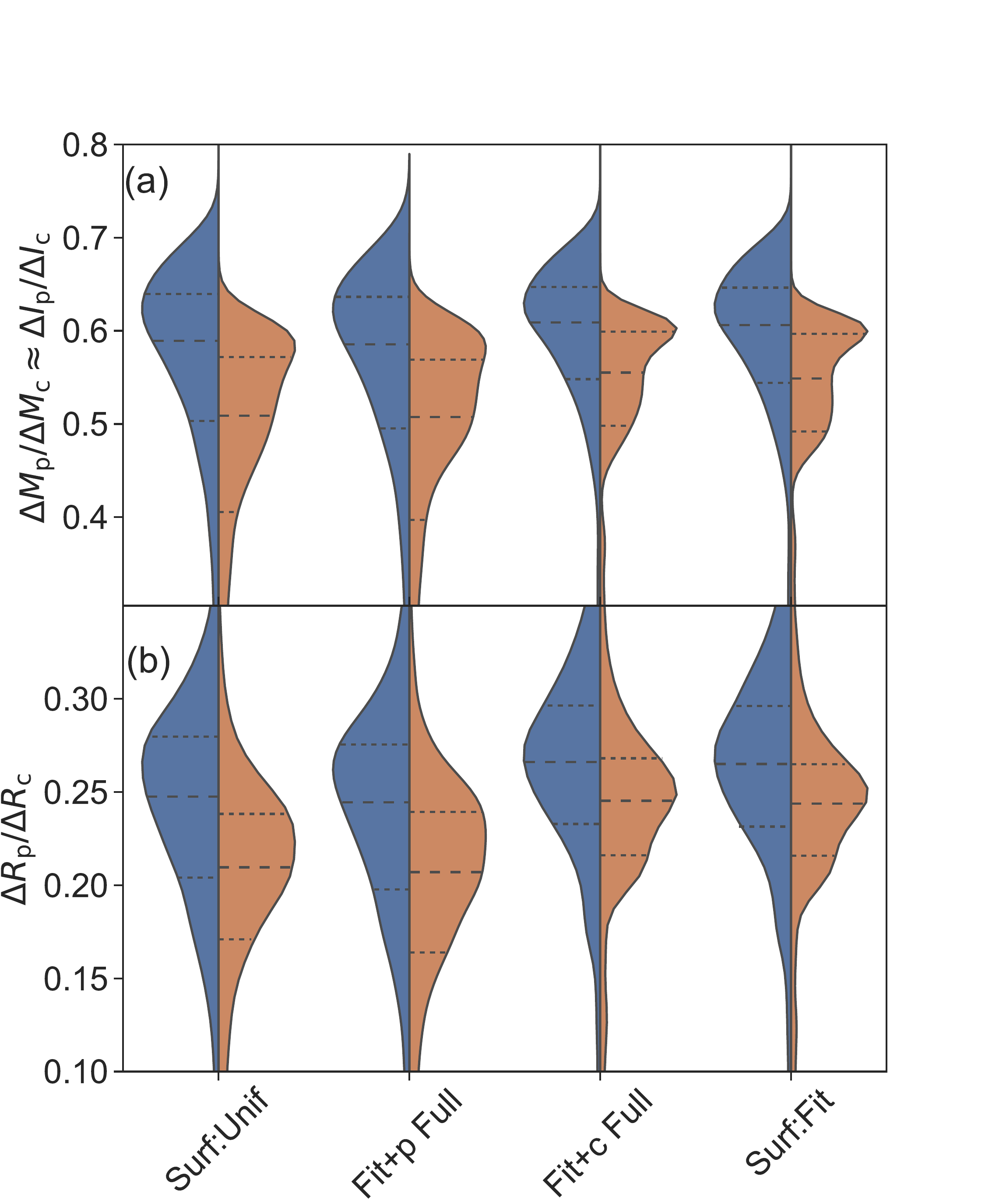}}
\caption{In order to determine the surface parameters that have the greatest effect on posteriors, we compare surface priors on $\Delta M_{\rm p} / \Delta M_{\rm c}$  (a) and $\Delta R_{\rm p} / \Delta R_{\rm c}$ (b), all obtained with the uniform symmetry energy prior \emph{Sym:Unif}. We show the \emph{Surf:Fit} prior, the distributions obtained by restricting all parameters to their fit values except $p$ (\emph{Fit+p:Full}), restricting all parameters to their fit values except $c$ (\emph{Fit+c:Full}) and finally the uniform surface prior \emph{Surf:Unif}. The results without data (blue) and the PNM posterior (orange) are shown.}  \label{fig:results}
\end{figure}

\begin{figure}[!th]
\vspace{-2.5cm}
\centerline{\includegraphics[scale=0.3]{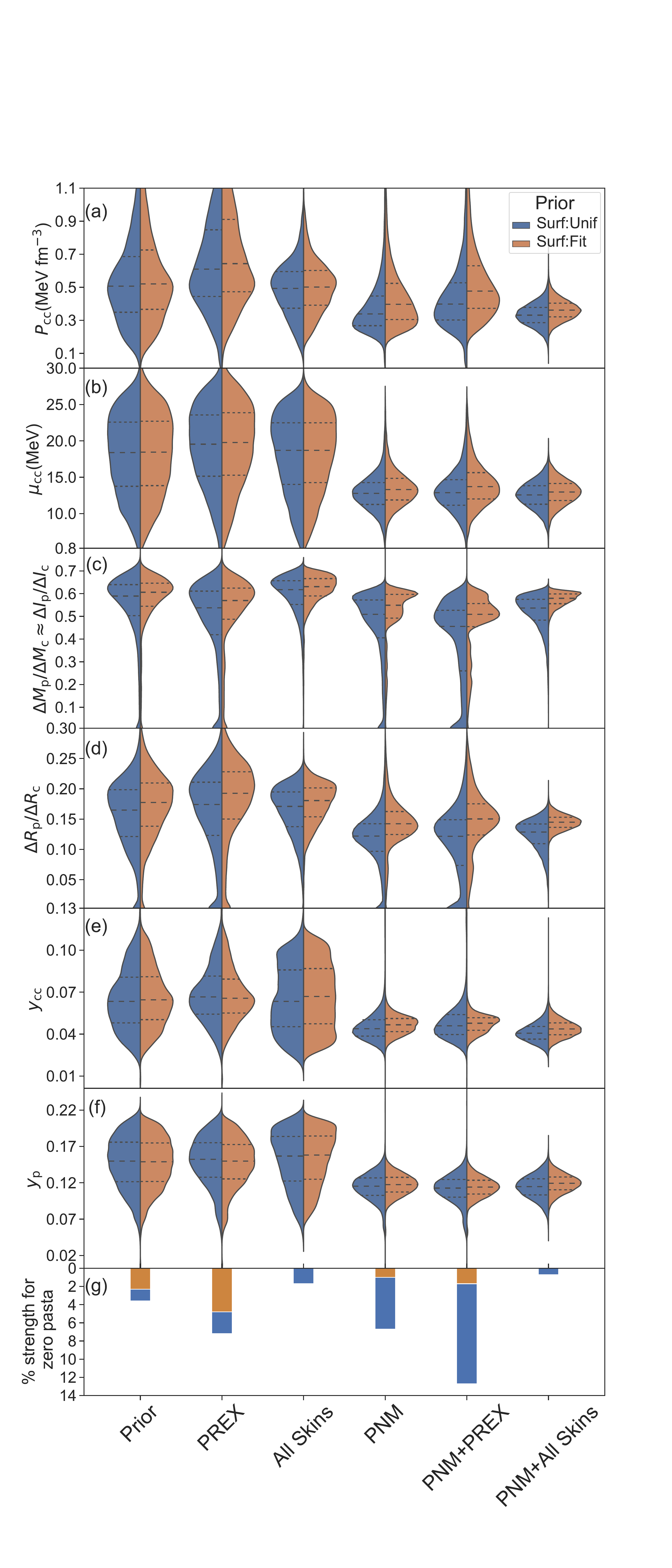}}
\vspace{-0.7cm}
\caption{Posterior distributions of the crust-core transition pressure (a), chemical potential (b), fractional mass/moment of inertia of pasta in the crust (c), fractional thickness of the pasta layer (d), crust-core transition proton fraction (e) and proton fraction at the transition to the pasta layers from spherical nuclei (f). The \emph{Sym:Unif} prior is used, together with the \emph{Surf:Unif} prior (blue), and \emph{Surf:Fit} prior (orange). Dashed and dotted lines show medians and 25th/75th percentiles respectively. The bars in the bottom panel show the percentage of models that predict no pasta.} \label{fig:results}
\end{figure}

\begin{figure}[!th]
\vspace{-0.8cm}
\centerline{\includegraphics[scale=0.35]{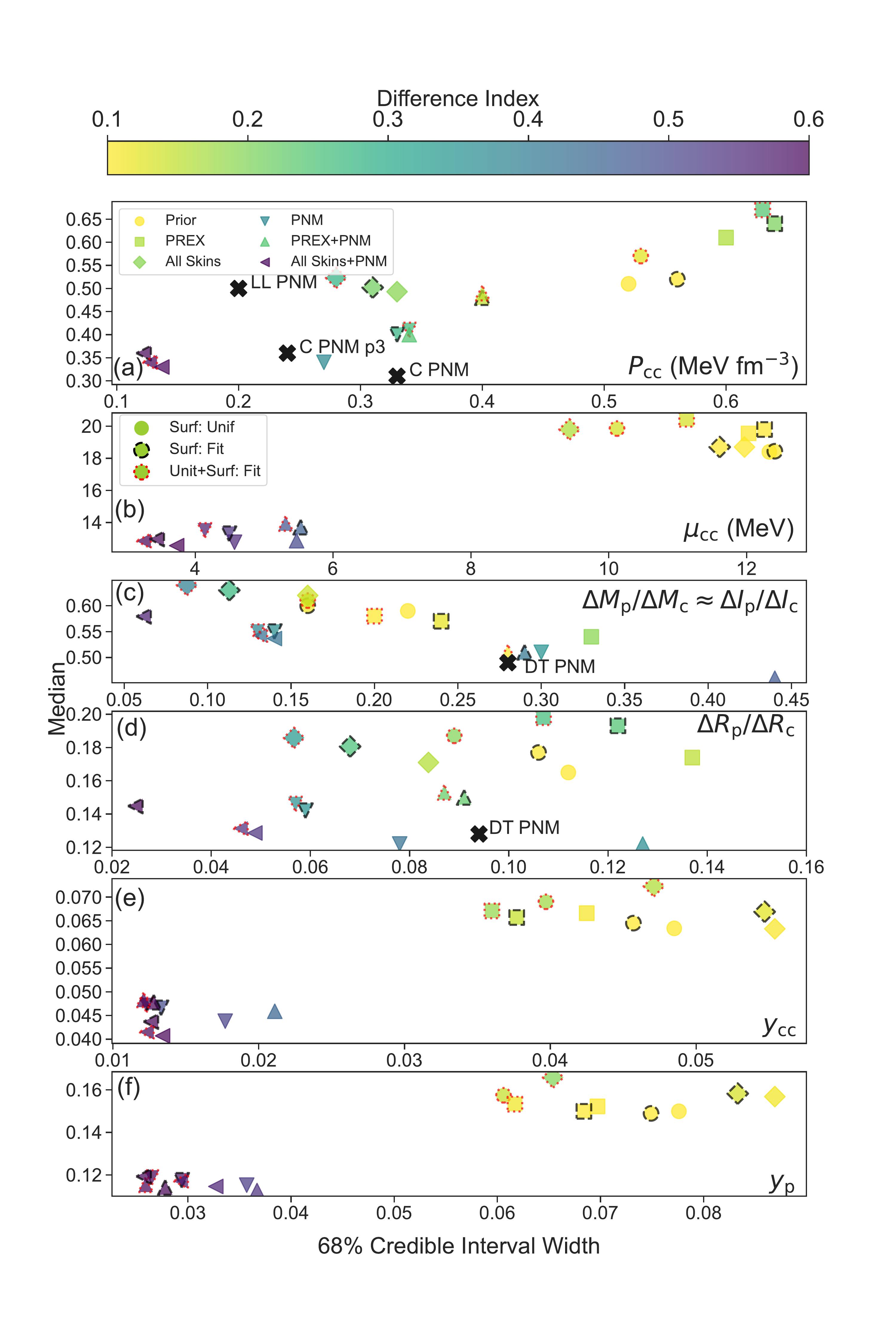}}
\vspace{-0.8cm}
\caption{Summary of all posterior distributions. The 68\% CI is plotted versus the median with the quantity and its units indicated at the right of each panel. Darker colors indicate larger difference indices (see text for details). \emph{Sym:Unif;Surf:Unif}, \emph{Sym:Unif;Surf:Fit}, and \emph{Sym:Unif+Unit;Surf:Fit} are indicated by shapes with no outline; black dashed outlines and red dotted outlines respectively. The priors alone are indicated by circles, the posteriors as different shapes as indicated in the legend. The crust-core transition pressure (a), chemical potential (b), fractional mass/moment of inertia of pasta in the crust (c), fractional thickness of the pasta layer (d), crust-core transition proton fraction (e) and proton fraction at the transition to the pasta layers from spherical nuclei (f) are shown. Comparison with other inferences of $P_{\rm cc}$ \cite{Lattimer:2013cr} (LL), \cite{Carreau:2019aa} (C) for transition pressure, and \cite{Dinh-Thi:2021lr} (DT) for mass/moment of inertia and thickness are shown, with the additional annotation indicating whether these other studies apply PNM constraints, and the range of $p$ used if applicable.}  \label{fig:results}
\end{figure}

%
%

\section{Results.}
\label{sec:results}

We first report on the symmetry energy constraints from the PREX measurement of $r^{208}_{\rm np}$. The 68\% credible intervals (CIs) are  $J=35.6^{+5.3}_{-6.6}$MeV, $L=83.2^{+31.5}_{-33.4}$MeV and $K_{\rm sym}=-149^{+199}_{-201}$MeV. The data does not constrain $K_{\rm sym}$ strongly. The $L$ parameter inferred is high $\approx50-114$MeV, but about $20$MeV smaller and with slightly smaller uncertainty than the value inferred from \cite{Reed:2021ux} from a more restricted range of models. The inferred value of $J$ is consistent with nuclear mass fits \cite{Lattimer:2013cr}.


We sample $\sim$100,000 points from our posteriors for each prior/dataset combination and calculate the resulting crust models.
In Fig.~3 we show the effect of the surface priors on the marginalized posterior distributions of $\Delta M_{\rm p} / \Delta M_{\rm c}$ (Fig.~3a) and $\Delta R_{\rm p} / \Delta R_{\rm c}$ (Fig.~3b), both calculated using the approximations outlined in Appendix A. We show the priors (blue) and the posteriors obtained from a representative dataset that gives significant constraints: PNM (orange). In each plot, from right to left, we show: our most restrictive surface prior, the fit to 3DHF calculations (the \emph{Surf:Fit} prior); the same except for allowing $c$ to assume its full range 2.0-7.0 (\emph{Fit+c Full}), the same but allowing $p$ to vary over its full range 2.0-4.0 (\emph{Fit+p Full}), and finally all surface parameters allowed to vary over their full ranges (the \emph{Surf:Unif} prior). 

The change in posterior from changing the prior from \emph{Surf:Unif} to \emph{Surf:Fit} is almost entirely due to relaxing the allowed range for $p$: the remaining surface parameters have much smaller effects. This is not surprising, as it is the parameter $p$ which controls the isospin dependence of the interface energy as the proton fraction approaches zero (see e.g. Fig.~5 of \cite{Gearheart:2011tg} and Fig.~1 of \cite{Carreau:2019aa}). The $P_{\rm cc}$ and $\mu_{\rm cc}$ distributions bear out the same conclusions.





In Fig.~4 we plot the prior and posterior distributions from the five datasets for the six crust quantities. Medians are represented by dashed lines and the distribution widths shown by dotted lines at the 25th and 75th percentiles. For each dataset, we show the distributions obtained using the \emph{Sym:Unif;Surf:Unif} prior (blue, left distributions) and the \emph{Sym:Unif;Surf:Fit} prior (orange, right distributions). 

In order to more clearly see trends in the posteriors, in Fig.~5 we plot the medians versus the 68\% credible limits for each quantity obtained with each combination of data and prior. The height of the plots are scaled by the relative variation in the medians across all datasets. The CIs are those with the median as the central point.  Each dataset and prior are represented by different markers and marker outlines respectively: no line (\emph{Sym:Unif; Surf:Unif}), dashed line (\emph{Sym:Unif; Surf:Fit}) and dotted line (\emph{Sym:Unif+Unit; Surf:Fit}). To track the data effects, compare different markers with the same border; to track prior effects, compare the same marker with different borders. 

To account for differences not captured by the median and CIs, we also calculate a difference index \cite{Pastore:2019} defined as $d(A,B)= 1-\int \textrm{min} [\rho_{\rm A}(x) \rho_{\rm B}(x)] dx$ where $x$ is the variable of interest, and $\rho_{\rm A,B}$ is the distribution of interest and the reference distribution (in this case the \emph{Sym:Unif; Surf:Unif} prior) respectively. The difference indices are indicated by the colors of the points in Fig.~5.

The inferred values of $P_{\rm cc}$, $\mu_{\rm cc}$, $\Delta M_{\rm p}/\Delta M_{\rm c}$, $\Delta R_{\rm p}/\Delta R_{\rm c}$, $y_{\rm cc}$ and $y_{\rm p}$ for each dataset are given in Table \ref{tab1}. The values inferred with the fewest initial modeling assumptions - \emph{Sym:Unif;Surf:Unif} priors - are given, with results from \emph{Sym:Unif; Surf:Fit} priors in parentheses. 


Finally, in Fig.~6 we show the two-dimensional posteriors for the \emph{Sym:Unif; Surf:Unif priors}. We show results for the prior, the PREX posterior and the Posterior with all our neutron skins data. The 1-$\sigma$ credible regions for $L$, $K_{\rm sym}$, $r^{208}_{\rm np}$, $P_{\rm cc}$, $\mu_{\rm cc}$ and $\Delta M_{\rm p} / \Delta M_{\rm c}$ and $\Delta R_{\rm p} / \Delta R_{\rm c}$ are displayed.



Comparing the prior (circular markers in Fig.~5), PREX (square markers) and Skins (diamond markers) datasets, the effect of the PNM-free data on the medians is to elevate them compare to the prior, except for the case of PREX data applied to $\Delta M_{\rm p} / \Delta M_{\rm c}$. The effect is relatively small, however. Then, adding the PNM data favors systematically smaller crusts, less pasta, and a smaller proton fraction throughout the pasta phases: medians with the PNM data included are about 25-30\% lower than without PNM data for all quantities except for $\Delta M_{\rm p} / \Delta M_{\rm c}$ which drops by $\approx$ 10\%.

As a result of the correlations with the neutron skin of $^{208}$Pb (see, e.g. Fig.~24 and the top panel of Fig.~27 in \cite{Balliet:2021lr}), tails in the posterior distributions develop at high values of $P_{\rm cc}$ and low values of $\Delta M_{\rm p} / \Delta M_{\rm c}$ and $\Delta R_{\rm p} / \Delta R_{\rm c}$, which leads to the PREX data \emph{increasing} the width of the CIs of  $P_{\rm cc}$, $\Delta M_{\rm p} / \Delta M_{\rm c}$ and $\Delta R_{\rm p} / \Delta R_{\rm c}$. Indeed, there is a smaller peak at zero pasta which can be understood as follows: as the pressures and chemical potentials of the pasta and crust-core transition approach each other, the amount of pasta shrinks. Then \emph{all} models that predict that the crust-core transition occurs first will predict zero pasta. The bars at the bottom of Fig.~4 show the fraction of models that predict no pasta in the crust for each dataset. The PREX data alone, the PNM data alone, and both combined give a significant strength for zero pasta $\sim 10\%$. The rest of the neutron skin data strongly favors pasta in the crust.

$P_{\rm cc}$, $\mu_{\rm cc}$,$y_{\rm cc}$ and $y_{\rm p}$ become significantly more constrained when we include the PNM data (the up, down and left triangles in Fig.~3), as can be seen by inspection in Fig.~4 and quantified in Fig.~5: the 68\% CIs shrink by 30-60\%. The PNM data do not systematically shrink the 68\% CIs for $\Delta M_{\rm p} / \Delta M_{\rm c}$ and $\Delta R_{\rm p} / \Delta R_{\rm c}$; although PNM data tightly constrains on $J$ and $L$, the correlation between $J$ and $L$ and the pasta parameters is relatively weak (see Fig.~22 and top panel of Fig.~27 in \cite{Balliet:2021lr} or Fig.~7 of \cite{Dinh-Thi:2021lr}).

As we add full neutron skin data (Prior$\to$Skins, PNM$\to$PNM+Skins) the 68\% CIs shrink by around 30-50\% for $P_{\rm cc}$ - so the neutron skin data is comparably informative to the PNM data for $P_{\rm cc}$ (therefore crust mass). The effect on the 68\% CIs for $\mu_{\rm cc}$, $y_{\rm cc}$ and $y_{\rm p}$ is around 10\%: PNM data is significantly more constraining for $\mu_{\rm cc}$ (therefore crust thickness), $y_{\rm cc}$ and $y_{\rm p}$. One can see in the top panel of Fig.~27 in \cite{Balliet:2021lr} that, starting from uniform priors, the quantities most correlated with $P_{\rm cc}$ are indeed the neutron skins and $L$ (the latter constrained by PNM data). We see this correlation in action in the two dimensional posteriors of Fig.~6: the full neutron skin dataset removes only a little strength at low values of $\mu_{\rm cc}$ which shows up in Fig.~4, whereas it removes more strength at both the low and high ends of the distributions of $P_{\rm cc}$, $\Delta M_{\rm p} / \Delta M_{\rm c}$ and $\Delta R_{\rm p} / \Delta R_{\rm c}$ resulting in tighter constraints.

\begin{figure}[!t]
\centerline{\includegraphics[scale=0.65]{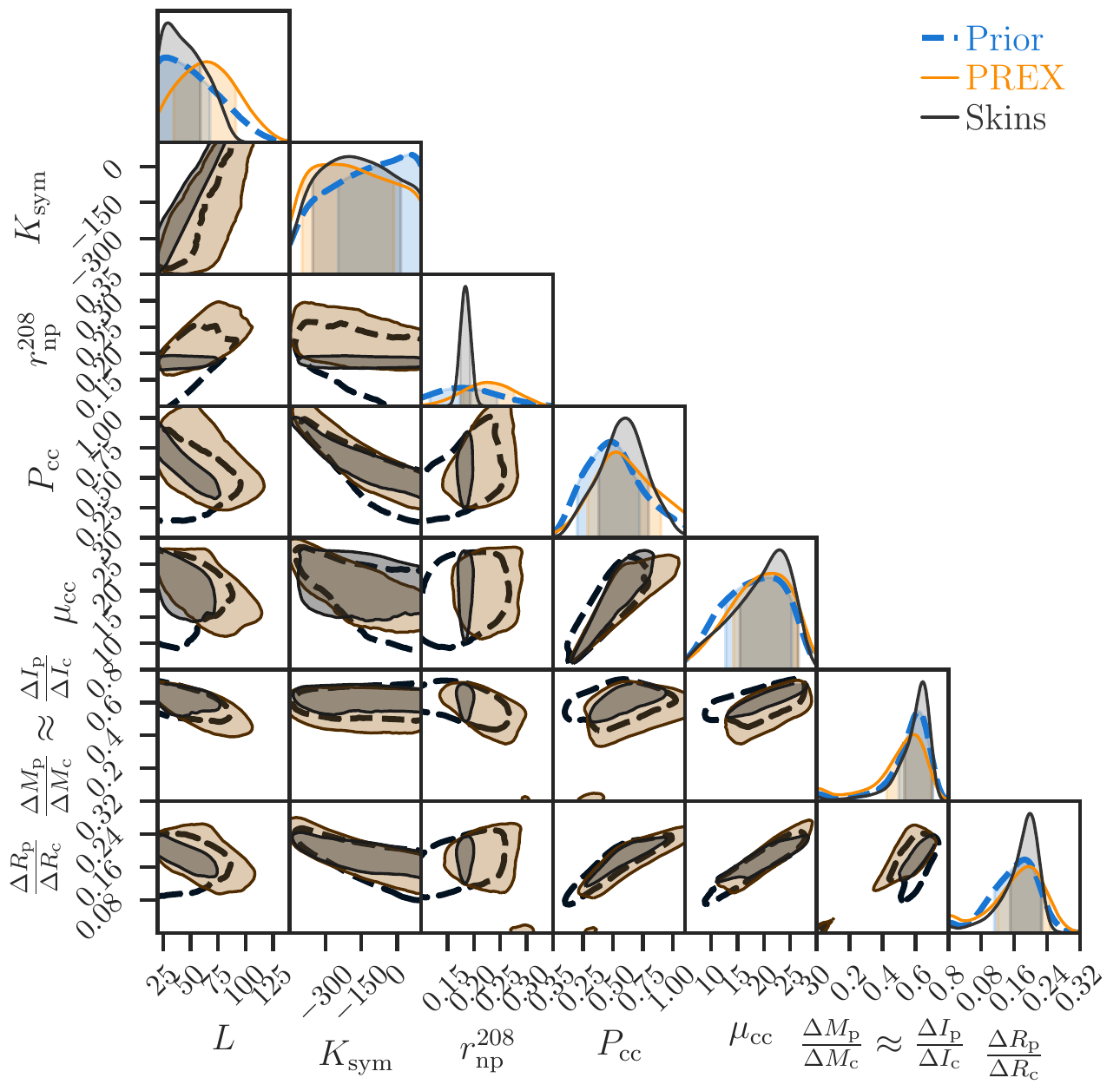}}
\caption{1-$\sigma$ contours of the two dimensional prior distribution (dashed line) and posterior distributions for PREX (light brown) and all neutron skin data (dark brown). Symmetry energy parameters $L$ and $K_{\rm sym}$, the neutron skin $r^{208}_{\rm np}$ and the crust quantities $P_{\rm cc}$, $\mu_{\rm cc}$, $\Delta M_{\rm p} / \Delta M_{\rm c}$ and $\Delta R_{\rm p} / \Delta R_{\rm c}$ are shown.}  \label{fig:results}
\end{figure}

\begin{table}[!t]
    \centering
    \caption{Table of median and 68\% (95\%) CIs for crust-core transition pressure, chemical potential,mass/moment of inertia,thickness fractions of pasta, crust-core proton fraction and proton fraction at the onset of pasta are shown using uniform surface priors (\emph{Surf:Fit} in parentheses) and the 5 different data combinations.}\label{tab1}
    \label{tab:model_params}
    \begin{tabular}{c|c|c}
        \hline
		Data & $P_{\rm cc}$ (MeV fm$^{-3}$) & $\mu_{\rm cc}$ (MeV) \\ 
		\hline
		Prior &  $0.51(0.52)^{+0.3(0.34)}_{-0.22(0.22)}$ & $18.4(18.4)^{+5.7(5.8)}_{-6.6(6.6)}$  \\
		
		Skins & $0.49(0.50)^{+0.15(0.18)}_{-0.15(0.16)}$ & $18.7(18.7)^{+5.1(5.1)}_{-6.8(6.5)}$  \\ 
		
		PREX & $0.61(0.64)^{+0.36(0.4)}_{-0.24(0.24)}$ & $19.6(19.8)^{+5.5(5.6)}_{-6.5(6.7)}$  \\ 
		
		PNM & $0.34(0.4)^{+0.17(0.21)}_{-0.10(0.12)}$ & $12.8(13.3)^{+2.3(2.4)}_{-2.3(2.1)}$  \\ 
		
		+Skins & $0.33(0.36)^{+0.07(0.06)}_{-0.07(0.06)}$ & $12.6(13.0)^{+1.8(1.7)}_{-1.9(1.7)}$ \\ 
		
		+PREX & $0.40(0.48)^{+0.21(0.25)}_{-0.13(0.15)}$ & $12.9(13.7)^{+2.8(3.1)}_{-2.7(2.5)}$ \\ 

		\hline \hline		
		
		 & $\Delta M_{\rm p}/\Delta M_{\rm c}\approx \Delta I_{\rm p} / \Delta I_{\rm c}$ & $\Delta R_{\rm p}/\Delta R_{\rm c}$  \\ 
		\hline
		Prior &   $0.59(0.61)^{+0.07(0.06)}_{-0.15(0.10)}$ & $0.165(0.177)^{+0.046(0.046)}_{-0.065(0.060)}$  \\
		
		Skins & $0.62(0.63)^{+0.05(0.05)}_{-0.11(0.065)}$ & $0.171(0.181)^{+0.033(0.028)}_{-0.051(0.040)}$  \\ 
		
		PREX & $0.54(0.57)^{+0.1(0.08)}_{-0.23(0.17)}$ & $0.17(0.19)^{+0.05(0.05)}_{-0.09(0.07)}$  \\ 
		
		PNM & $0.51(0.55)^{+0.08(0.06)}_{-0.22(0.08)}$ & $0.122(0.142)^{+0.031(0.032)}_{-0.047(0.027)}$ \\ 
		
		+Skins & $0.54(0.58)^{+0.05(0.03)}_{-0.09(0.03)}$ & $0.129(0.145)^{+0.019(0.012)}_{-0.030(0.014)}$ \\ 
		
		+PREX & $0.46(0.51)^{+0.10(0.07)}_{-0.34(0.22)}$ & $0.122(0.150)^{+0.041(0.041)}_{-0.086(0.050)}$  \\
		
		\hline \hline
		
		& $y_{\rm cc}$ & $y_{\rm p}$ \\ 
		\hline
		Prior & 
		$0.063(0.065)^{+0.027(0.025)}_{-0.022(0.021)}$
		&  
		$0.145(0.149)^{+0.036(0.036)}_{-0.042(0.039)}$ \\
		
		Skins &  $0.063(0.067)^{+0.031(0.028)}_{-0.028(0.027)}$ & $0.157(0.158)^{+0.036(0.035)}_{-0.051(0.049)}$ \\ 
		
		PREX & $0.067(0.066)^{+0.024(0.022)}_{-0.018(0.016)}$ & $0.152(0.15)^{+0.032(0.032)}_{-0.037(0.036)}$ \\ 
		
		PNM &  $0.044(0.047)^{+0.01(0.006)}_{-0.008(0.007)}$ & $0.115(0.118)^{+0.017(0.015)}_{-0.019(0.015)}$ \\ 
		
		+Skins & $0.041(0.044)^{+0.007(0.007)}_{-0.006(0.006)}$ & $0.115(0.119)^{+0.016(0.013)}_{-0.017(0.013)}$\\ 
		
		+PREX & $0.046(0.048)^{+0.012(0.006)}_{-0.009(0.007)}$ & $0.111(0.116)^{+0.017(0.013)}_{-0.017(0.014)}$ \\
		\hline

    \end{tabular}
\end{table}


In Fig.~4, the shapes of the \emph{Surf:Unif} and \emph{Surf:Fit} priors can be seen to be broadly similar for all datasets, and in Fig.~5 markers of the same shape (data) cluster more than those with the same border (prior), and that changes in the posterior's shape is driven by changing data (compare colors between priors and between data). Note that in Fig.~5 we also include the points corresponding to the unitary gas prior \emph{Sym:Unif+Unit} which makes only a small difference to the median values. As might be expected, the \emph{Surf:Fit} prior and the \emph{Sym:Unif+Unit} prior, which incorporate more information into the priors, have the effect of shrinking the 68\% CI. For $P_{\rm cc}$, $\mu_{\rm cc}$, $y_{\rm cc}$ and $y_{\rm p}$, the effect of the \emph{Surf:Fit} prior compared to the \emph{Surf:Unif} prior is no more than that of the data. Additionally, it is clear that the main trend is for greater difference index (color in Fig.~5) to correlate with smaller CI, and that the data gives the dominant effect. 

The exception to this is $\Delta M_{\rm p} / \Delta M_{\rm c}$ and $\Delta R_{\rm p} / \Delta R_{\rm c}$ in the case when we include PNM and all skins data; then \emph{Surf:Fit} priors result in smaller CIs than \emph{Surf:Unif} priors by over 50\%. In this case, the priors on the surface energy parameters are more informative than the data. This highlights the importance of finding more robust constraints on the surface energy (noted also in \cite{Dinh-Thi:2021lr}). Additionally, the difference index does not appear to be correlated particularly with the change in CI for these two quantities: the data and priors mainly change the shape of the distribution characterized by its higher moments - for example, the development of a second peak at zero pasta.

%
%

\section{Summary and Conclusions.} 
\label{sec:summary}

The combination of PNM and neutron skin data gives the tightest constraints, and are most robust when uniform priors are used. A significant amount of pasta - 50\% by mass and moment of inertia, 12\% by thickness - is strongly favored, although the possibility that no pasta exists in the crust cannot be ruled out; 12\% of models predict pasta is absent for the PREX+PNM dataset. The median proton fraction in the pasta phases is predicted to decrease from $\approx$16\% at the top of the pasta layer to $\approx$7\% at the crust-core transition when only neutron skin data is taken into account, and from $\approx$12\% down to $\approx$4.5\% when PNM data is added. This is of relevance to, for example, possible direct Urca processes in the pasta layers that may enhance the cooling of the neutron star \cite{Newton:2013dz}. We note here that extending the chiral-EFT analysis of \cite{Drischler:2020ab} to infer $K_{\rm sym}$ should lead to even tighter constraints on the crust.


Although we have attempted a thorough characterization of the uncertainties within our CLDM, there are certainly additional uncertainties we should work towards quantifying. Although our extended EDF allows a wide parameter space exploration of the symmetry energy, the third order symmetry energy parameter, $Q_{\rm sym}$, also correlates with the crust parameters  \cite{Carreau:2019aa,Dinh-Thi:2021lr}. The surface energy function is not unique, and extensions and alternatives have been studied (see e.g. \cite{Watanabe:2000yq,Lorenz:1991thesis,Newton:2013sp,Carreau:2019aa}). More microscopic alternatives to the CLDM such as Thomas-Fermi \cite{Bao:2015kx} and Hartree-Fock \cite{Newton:2022lr} methods give results that appear consistent with the CLDM, but a more thorough comparison should be done. 

Nevertheless, there is a concordance between our results and similar recent statistical predictions, as can be seen by the black crosses in Fig~4. Constraints obtained using a similar CLDM \cite{Carreau:2019aa,Dinh-Thi:2021lr} constrained by low density PNM calculations are shown. For $P_{\rm cc}$, a range of $p=$2.5-3.5 (`C PNM') and a single value of $p=3$ (`C PNM p3') were used, so these are most comparable with \emph{Surf:Unif} and \emph{Surf:Fit} priors respectively. For $\Delta R_{\rm p}/\Delta R_{\rm c}$ and $\Delta M_{\rm p}/\Delta M_{\rm c}$ the same model with surface parameters fit to nuclear masses \cite{Dinh-Thi:2021lr} (`DT PNM') was used. The above studies used a more schematic EDF that allowed for $Q_{\rm sym}$ to be varied too. Also shown it the constraint on $P_{\rm cc}$ from \cite{Lattimer:2013cr} (`LL PNM') obtained by examining the stability of nuclear matter with respect to small density fluctuations, and was also constrained by PNM calculations (but imposes a correlation between $J,L$ and $K_{\rm sym}$). In all cases, the best comparison with our calculations is with the `PNM' dataset (triangles). Let us also note the classic work of \cite{Lorenz:1993qy} predicted a relative mass and moment of inertia of pasta of around 0.5. Previous studies have not included neutron skin data. The thermodynamic spinodal method of \cite{Lattimer:2013cr} is known to give a upper bound on the transition point. Given this, the models are in remarkable agreement. We can conclude that we are entering an era where statistically meaningful constraints on neutron star crusts from experimental and theoretical data are becoming competitive with modeling uncertainties.

Our results point to there being a large amount of pasta in the neutron star crust, which has a number of notable observational consequences. If disordered, as microscopic simulations of pasta suggest \cite{Newton:2022lr}, this predicts that the cooling timescale of the neutron star crust will be significantly enhanced at late times. It also suggests that magnetic fields in the crust would decay rapidly in pulsars with ages $\sim10^5-10^6$ years \cite{Pons:2013ly}.
We should expect the shear modulus at the crust-core interface to be modified by the existence of pasta, which has implications for a number of scenarios. If the shear modulus is smaller, for example:  pasta could less efficiently damp r-modes in rotating neutron stars, therefore allowing greater spin-up of millisecond pulsars \citep{Wen:2012aa,Vidana:2012aa}; pasta would decrease the resonant frequency of crust $i$-modes possibly observable in resonant shattering of the crust just prior to two neutron stars merging \cite{Neill:2021tg}; the maximum size of mountains, and hence their detectability as persistent gravitational wave sources, would be reduced \cite{Gearheart:2011tg}. A large amount of pasta increases the direct Urca emissivity of the star \cite{Gusakov:2004fk} resulting in enhanced cooling early on and subsequently decreasing the rapid cooling rate brought on by the onset of superfluidity in young neutron stars \cite{Newton:2013dz}. The effects of pasta on the dynamics of neutron vortices in the inner crust, a key ingredient of many glitch models, have yet to be modeled. Therefore our work suggests renewed efforts to understand the microphysics of nuclear pasta are required.

\section{Acknowledgements}
This work is supported in part by the Physics and Astronomy Scholarship for Success (PASS) project funded by the NSF under grant No. 1643567 and the NASA grant 80NSSC18K1019.

\appendix

\section{Approximations for mass and thickness of pasta.--}
\label{sec:sample:appendix}

Our starting point is equations~27 and 28 from
\cite{Lattimer:2007lr}. We define $x_i=2(\mu_i-\mu_0)/m_{\rm b }c^2$ where $i={\rm cc,p}$ labels the location at which to calculate the quantities, the crust-core transition and onset of pasta respectively, and $\mu_0 \approx 9$ MeV is the baryon chemical potential at the surface of the star. Then we define

\begin{equation}
    \mathcal{G}_i=e^{x_i},
\end{equation}

\noindent from which the thickness relative to the stellar radius $R$ of the whole crust $i={\rm cc}$ or the crust above the pasta phases $i={\rm p}$ is given as

\begin{equation}
    \frac{\Delta R_i}{R} =\frac{\mathcal{G}_i-1}{\mathcal{G}_i(1-2\beta)^{-1}-1},
\end{equation}

\noindent where $\beta=GM/Rc^2$ is the stellar compactness. Now, $x_i \ll 1$ so $\mathcal{G}_i \approx 1 + x_i$ and $\mathcal{G}_i - 1 \ll 1 $.

For most neutron stars $\beta < 0.2$ and so $(1-2\beta)^{-1} \approx 1 + 2\beta$, and the denominators read

\begin{align}
    \mathcal{G}_i(1-2\beta)^{-1}-1 &\approx \mathcal{G}_i - 1 + 2 \mathcal{G}_i \beta \notag \\ &\approx 2 \mathcal{G}_i \beta = 2(1+x_i)\beta \approx 2\beta,
\end{align}

\noindent and

\begin{equation}
    \frac{\Delta R_i}{R} \approx \frac{x_i}{2\beta}.
\end{equation}

The thickness of the pasta phases is 

\begin{equation}
    \frac{\Delta R_{\rm p}}{R} = \frac{\Delta R_{\rm cc}-\Delta R_{\rm p}}{R} \approx \frac{x_{\rm cc} - x_{\rm p}}{2\beta},
\end{equation}

\noindent so the relative thickness of the pasta layers to the crust is

\begin{equation}
\frac{\Delta R_{\rm p}}{\Delta R_{\rm c}} \approx \frac{x_{\rm cc} - x_{\rm p}}{x_{\rm cc}} =
\frac{\mu_{\rm cc} - \mu_{\rm p}}{\mu_{\rm cc}-\mu_0}.
\end{equation}

Similar expressions were derived in \cite{Zdunik:2016fk}, where it was demonstrated that the approximations made led to an accuracy of $<1\%$ in the crust thickness, so similar accuracy can be expected for equation~A.6.

The mass of a layer in the crust is proportional to the pressure at the bottom of the layer \cite{Lorenz:1993qy,Zdunik:2016fk}, so more straightforwardly

\begin{equation}
\frac{\Delta M_{\rm p}}{\Delta M_{\rm c}} \approx \frac{P_{\rm p}}{P_{\rm cc}}. 
\end{equation}

As shown by Eq.~3 in \cite{Lorenz:1993qy} the moment of inertia of the crust is proportional to the mass of the crust to a first order of approximation, and so $P_{\rm cc}$ controls the moment of inertia of the crust too, and $\Delta M_{\rm p} / \Delta M_{\rm c}\approx \Delta I_{\rm p} / \Delta I_{\rm c}$. This is supported by inspecting Table.~I of \cite{Lorenz:1993qy} and Table~5 and Fig.~8 of \cite{Dinh-Thi:2021lr}.

\bibliographystyle{elsarticle-num} 





\end{document}